\documentclass[superscriptaddress,twocolumn,showpacs]{revtex4}

\usepackage{graphics}% Include figure files
\begin{document}

%%%%%%%%%%%%%%%%%%%%%%%%%%%%%%

% Include the paper's title here

\title{Micro-bias and macro-performance}

\author{S. M. D. Seaver} 

\affiliation{Department of Chemical and Biological
Engineering,Northwestern University, Evanston, IL 60208, USA}

\affiliation{Northwestern Institute on Complex Systems, Northwestern
  University, Evanston, IL, 60208, USA}

\author{A. A. Moreira}

\affiliation{Departamento de F\'{\i}sica, Universidade Federal do
Cear\'{a}, Cear\'{a}, Brazil}

\author{M. Sales-Pardo}

\affiliation{Department of Chemical and Biological
Engineering,Northwestern University, Evanston, IL 60208, USA}

\affiliation{Northwestern Institute on Complex Systems, Northwestern
  University, Evanston, IL, 60208, USA}

\affiliation{Northwestern University Clinical and Translational
  Sciences Institute, Chicago, IL 60611, USA}

\author{R. D. Malmgren}

\affiliation{Department of Chemical and Biological
Engineering,Northwestern University, Evanston, IL 60208, USA}

\affiliation{Northwestern Institute on Complex Systems, Northwestern
  University, Evanston, IL, 60208, USA}

\author{D. Diermeier}

\affiliation{Managerial Economics and Decision Sciences, Kellogg School of
Management, Northwestern University, Evanston, IL 60208, USA}

\affiliation{Northwestern Institute on Complex Systems, Northwestern
  University, Evanston, IL, 60208, USA}

\author{L. A. N. Amaral}

\affiliation{Department of Chemical and Biological
Engineering,Northwestern University, Evanston, IL 60208, USA}

\affiliation{Northwestern Institute on Complex Systems, Northwestern
  University, Evanston, IL, 60208, USA}

\email{amaral@northwestern.edu}

\date{Received: 1 August 2008}
%% The \maketitle command is necessary to build the title page.

%%%%%%%%%%%%%%%%%%%%%%%%%%%%%%%%%%%%%%%%%%%%%%%%%%%%%%%%%%%%
%%%%%%%%%%%%%%%%%%% ABSTRACT %%%%%%%%%%%%%%%%%%%%%%%%%%%%%%%
%%%%%%%%%%%%%%%%%%%%%%%%%%%%%%%%%%%%%%%%%%%%%%%%%%%%%%%%%%%%
\begin{abstract}
  We use agent-based modeling to investigate the effect of
  conservatism and partisanship on the efficiency with which large
  populations solve the density classification task---a paradigmatic
  problem for information aggregation and consensus building. We find
  that conservative agents enhance the populations' ability to
  efficiently solve the density classification task despite large
  levels of noise in the system. In contrast, we find that the
  presence of even a small fraction of partisans holding the minority
  position will result in deadlock or a consensus on an incorrect
  answer. Our results provide a possible explanation for the emergence
  of conservatism and suggest that even low levels of partisanship can
  lead to significant social costs.
\end{abstract}

\pacs{ {87.23.Ge} {Dynamics of social systems}, {89.75.-k} {Complex systems }}

\maketitle

%%%%%%%%%%%%%%%%%%%%%%%%%%%%%%%%%%%%%%%%%%%%%%%%%%%%%%%%%%%%
%%%%%%%%%%%%%%%%%%% BODY OF THE PAPER %%%%%%%%%%%%%%%%%%%%%%
%%%%%%%%%%%%%%%%%%%%%%%%%%%%%%%%%%%%%%%%%%%%%%%%%%%%%%%%%%%%
%%%%%%%%%%%%%%%%%%%%%% INTRODUCTION %%%%%%%%%%%%%%%%%%%%%%%%
\section{Introduction}

Many practical and scientific problems require the collaboration of
groups of experts, with different expertise and background.
Remarkably, it turns out that large groups of cooperative agents are
extremely adept at finding efficient strategies for solving such
problems~\cite{gigerenzer99,gigerenzer00}; the development of the
scientific method within the physical science or the development of
entire suites of computer software by the open source movement
~\cite{goldman05} being perhaps two of the most notable
instances~\cite{giere05}. Indeed, even loosely structured groups have
demonstrated an ability to coordinate and find innovative solutions to
complex problems.

In the corporate world, several companies---including IBM, HP and
various consulting companies---have used ``the wisdom of the crowd''
principle as the justification for the creation of knowledge
communities, the so-called ``Communities of Practice,'' which have
enabled organizations to spawn new ideas for products and
services~\cite{lesser01}. Other companies, such as Intel, Eli Lilly,
and Procter \& Gamble, which created company-sponsored closed knowledge
networks, are now opening them to outsiders \cite{chesbrough03}.

Recognizing that knowledge exists not merely in the members of the
network but in the networks themselves---that is, in the members {\it
and\/} in their interactions---naturally leads to the question of what
characterizes successful communities and what measures could be taken
to improve the ability of groups and organizations to
innovate. Previous work investigated the importance of group
diversity~\cite{page01,march91}, team formation~\cite{guimera05c}, and
the structure of the interaction
network~\cite{castellano03,moreira04,sood05,uzzi05,suchecki05}.  Here,
we focus instead on the effect of micro-level strategies on
macro-level performance.

Recent work demonstrates that under quite general conditions,
well-intentioned and completely trusting agents can efficiently solve
information aggregation and coordination tasks \cite{moreira04}.  We
investigate how changes in the intention and trust level of agents
affects the efficiency of solving such tasks by considering three
types of agents: naives, conservatives and partisans. Naive agents are
well-intentioned and completely trusting. Conservative agents are
well-intentioned but not completely trusting. Partisan agents are
neither well-intentioned nor completely trusting.

Remarkably, we find that conservative agents, despite slowing the
information aggregation and coordination process, actually enhance the
populations' ability to efficiently solve these tasks under large
levels of noise. In contrast, we find that even a small fraction of
partisans holding the minority position will result in deadlock or a
consensus on an incorrect answer. Significantly, only by completely
disregarding partisan opinions, can the population recover its
original ability to solve the task.

%----------- METHODS ------------
\section{The Model}

We use here the density classification task, a model of decentralized
collective problem-solving~\cite{crutchfield95}, to quantitatively
investigate information aggregation and coordination. The density
classification task is completed successfully if (i) all agents
converge to the same state within a determined time period, and (ii)
that consensual state was the majority state in the initial
configuration.

Before proceeding, let us explain the reasons why the density
classification task is a good paradigm for the type of problems into
which we aim to gain insight. Consider a population of agents tackling
a problem in which there is a large uncertainty and for which no agent
will be able, by herself, to demonstrate that a particular solution is
correct. If one assumes that all agents are well-intentioned, that is,
that they want to find a good solution to the problem, then it is
plausible to assume that the answer reached by a specific agent
``contains'' a good answer distorted by some noise.  Under these
conditions, an efficient strategy is to aggregate the answers from all
agents, as information aggregation cancels the distorting component of
the individual answers. However, in many situations a centralized
structure may not be practical or desirable because it is too
inefficient, too costly, or because it would be difficult to secure an
unbiased central authority. For these reasons, decentralized
strategies may be preferable or even the only ones feasible.
\\

%----------- Description of agents ------------
\subsection{The Agents }

We consider a population of $N$ agents. For simplicity, and without
loss of generality. The state of each agent is a binary variable $a_j
\in \{-1,1\}$ that represents the answer to a stated problem. Updating
occurs using Boolean functions. As we indicated earlier, agents can
hold various types of intentions. Specifically, well-intentioned
agents have a bias $b_j$ toward their present state $b_j = a_j(t)$,
while ``partisan'' agents have a bias toward a particular state, for
example, $b_j = -1$. Naive and conservative agents are thus not biased
for or against either state {\it per se}, they merely prefer whatever
answer they currently hold. While all three types of agents may change
their state in response to peer pressure, a partisan agent will defect
back to his preferred state if peer pressure decreases below a
threshold value.

Both conservative and partisan agents can have different levels of
trust on their neighbors, that is, different thresholds for responding
to peer pressure. We define the ``strength'' $s_j\in[0,1]$ of agent
$j$'s ``conviction'' as the threshold that must be exceed by
$\Delta_j(t)$---which is the difference between the fraction of
majority and minority positions among agent $j$'s neighbors---for
agent $j$ to change states. If $s_j=0$, then the agent is completely
trusting, that is, naive, whereas if $s_j=1$, the agent is completely
distrusting and thus will never change his answer (Fig. 1).

Formally, one can write the update rule for agents with naive,
conservative, or partisan strategies as:
\begin{eqnarray}
a_{j}(t + 1) = \left\{
    \begin{array}{rrl}
      +1 & & \Delta_{j}(t) > -b_{j}s_{j}\\
       a_{j}(t) & & \Delta_{j}(t) = -b_{j}s_{j} \,,\\
      -1 & & \Delta_{j}(t) < -b_{j}s_{j}
    \end{array}
    \right.
\label{eqn:par_maj}
\end{eqnarray}
\noindent
where $\Delta_{j}(t)$ is defined as:
\begin{eqnarray}
\Delta_{j}(t) = \frac{1}{1+k_{j}} \left( a_{j}(t) + \sum_{l=1}^{k_{j}}
\widetilde a_{l}^{j}(t) \right).
\label{eqn:ave_state}
\end{eqnarray}
\noindent
Here, $k_j$ is the number of neighbors of agent $j$ and $\widetilde
a_{l}^{j}(t)$ is the perceived state of neighbor $l$ by agent $j$ at
time $t$, which may differ from $a_l(t)$ due to
noise~\cite{moreira04}. We implement the effect of noise by picking
with probability $\eta$ a value for $\widetilde a_{l}^{j}(t)$ that is
drawn with equal probability from \{$-1,1$\}.  If $\eta=0$, then
$\widetilde a_{l}^{j}(t)=a_l(t)$, whereas for $\eta=1$, $\widetilde
a_{l}^{j}(t)$ is a random variable. Figure~1 illustrates the response
of each type of agent to different signals.

The model we study is, on the surface, quite similar to the voter
model which has been widely used to study social dynamics and opinion
formation~\cite{castellano03,sood05,suchecki05}. A significant
difference, however, is that whereas in the voter model the agent
picks a single neighbor at random and adopts its neighbor's state, the
model we use is more similar to that of an Ising model with zero
temperature Glauber dynamics, in which at each step the agent tries to
align with some local field exerted by the neighbors and herself. This
subtle difference leads the two models to quite distinct dynamics.

Tessone and Toral~\cite{tessone08} have recently studied opinion
formation in a model in which agents are have preference toward a
specific opinion with a variety of strengths. Each agent follows the
simple majority of their neighbors to update her state taking into
account its bias. The authors find that the system responds more
efficiently to external forcing if the agents are diverse, that is
each agent has a different bias strength. In here, however, we focus
on the effect of opinion-bias on consensus formation without external
forcing.

\subsection{Network Topology }

A large body of literature demonstrates that social networks have
complex topologies~\cite{girvan02}, and yet have common
features~\cite{guimera03,newman04b}. We build a network following the
model proposed by Watts and Strogatz~\cite{watts98,watts99}, which,
despite its simplicity, captures two important properties of social
networks: local cliquishness and the small-world property. We
implement the Watts and Strogatz model as follows. First, we create an
ordered network by placing the agents on the nodes of a
one-dimensional lattice with periodic boundary conditions.  Then, we
connect each agent to its $k$ nearest neighbors in the lattice.  Next,
with probability $p$, we rewire each of the links in the network by
redirecting a link to a randomly selected agent in the lattice.  By
varying the value of $p$, the network topology changes from the
ordered one-dimensional lattice ($p$ = 0) to a random graph ($p$ =
1). We verified that our results are robust to changes in $p$ as long
as the network has a small-world topology, which for $N=401$ occurs
for $p\geq0.1$ (Supporting Online Material). In this study, we
investigate populations of $N=401$ agents placed on a one-dimensional
ring lattice where each agent has $k=6$ neighbors.  To implement a
small-world topology we rewire each connection with probability
$p=0.15$.  Recent research demonstrates that, under these conditions,
the naive heuristic enables the efficient convergence of the system to
the correct consensus ~\cite{moreira04,hastie05}. This finding is
similar to what has been reported for the voter model. Specifically,
studies of the voter model on complex topologies have shown that
finite systems convergence faster to a consensus in small-world
networks than in regular lattices in one dimension and that this
effect is independent of the degree
distribution~\cite{castellano03,sood05,suchecki05}.

\section{Results}

\subsection{Effect of conservatism }

We first consider the effect of conservative agents on the efficiency
of the system. Let's assume that the system has both naive and
conservative agents present, and that the fraction of agents using
conservative strategies is $f_c$.  The characteristics of the
population are then described by the distribution
\begin{equation}
P(b_j,s_j) =  \delta_{b_j,a_j(t)}\;P_s(s_j)\,.
\label{eqn:5}
\end{equation}
\noindent
For simplicity, we set $s_j=s>0$ for the conservative agents and $s_j=0$
for the naive agents.  Thus,
\begin{equation}
P_s(s_j) = (1-f_c) \;\delta(s_j-0)+f_c\; \delta(s_j-s).
\label{eqn:6}
\end{equation}

We study three values for the bias strength: $s=2/7$, $4/7$, and $6/7$
(Fig. 2A).  For $s=2/7$ and $s=4/7$, the system completes the density
classification task with extraordinary efficiency. Indeed, increasing
$f_c$ results in greater efficiencies for high noise levels,
which can be explained if one considers the stabilizing effect of
conservative agents on the dynamics.

In order to further characterize the effect of conservative agents on
the system's efficiency, we next investigate how the time needed for
the system to reach the steady state depends on $f_c$ (Fig.~3
and Supporting Online Material).  We find that the ``convergence
time'' grows quite rapidly with the fraction of conservatives in the
system.  In particular, for $f_c>0.3$ the system can no longer reach
the steady state within $2N$ time steps.

These two findings suggest that a population of agents trying to
optimize strategies in order to reach maximum efficiency must balance
the greater accuracy of the system in completing the task for larger
noise levels afforded by larger fractions of conservative agents with
the rapidly increasing convergence time as $f_c$ increases.

\subsection{Effect of partisanship }

We next consider the effect of partisan agents on the efficiency of
the system. Let's assume that the system now has both naive and
partisan agents present, and that the fraction of agents using
partisan strategies is $f_p$. Because partisan agents can have bias
toward two distinct answers, we consider two scenarios. In the first
scenario, all partisan agents have a bias toward ``$-1$'', that is,
they prefer the incorrect answer to the density classification task
the population is trying to complete ~\footnote{If all partisan agents
have a bias toward ``$1$'', because the preferred state matches the
majority state, the efficiency is high for any value of $\eta$, $f_p$,
and $s$}.  If $s_j=s>0$, then the characteristics of the population in
this scenario are described by the distribution
\begin{equation}
  P(b_j,s_j) = (1-f_p) \delta(s_j-0)   \delta_{b_j,a_j(t)} +f_p \delta(s_j-s)  \delta_{b_j,-1}\,.
\label{eqn:7}
\end{equation}
\noindent
We again study three values for the bias strength: $s=2/7$, $4/7$, and
$6/7$.  Our results reveal that even when partisan agents have a small
bias strength $s=2/7$, and, therefore, yield to peer pressure easily,
 $f_p\ge 0.15$ is enough to overcome the initial majority and
lead the population to converge to the incorrect answer (Fig. 2B).

In the second scenario, we consider a balanced distribution of
partisans with $f_p/2$ agents partisan toward answer ``$1$'' and
$f_p/2$ agents partisan toward answer ``$-1$'' (Fig. 2C).  For
$\eta<0.4$ and $f_p<0.3$, we find that the strength of the initial
majority is able to drive the population to a consensus on the correct
answer. However, if either the noise level or the fraction of
partisans increases, the population settles into a
deadlock. One could naively expect that in the high bias
strength case the results for partisans and conservatives should be
identical. In both systems, conservatives and partisans alike are
frozen in their preferred state. In the system with conservatives, the
distribution of preferred states is exactly the same as the
distribution of initial states, which is that 57\% of agents prefer
state ``$1$''. However, in the system with partisans, the distribution
of preferred states is such that 50\% of the agents prefer state
``$1$''. Therefore, even though the distribution of initial states is
57\%, all partisans will switch to their preferred state because of
the high bias strength. As a consequence, in the regions of lowest
efficiency, there is a difference in efficiency between the results
for conservatives and the results for partisans.

The effect of strongly biased partisans in opinion formation, that is
those partisans that never change opinion, has also been studied in
the context of the voter model~\cite{mobilia03,mobilia07} showing that
only one partisan is enough to significantly slow down consensus
formation and that, in regular lattices and complete graphs, when
an equal number of partisans $P$ of each opinion is present, the
efficiency of the system in the steady state is Gaussian distributed
with zero mean and variance $\propto 1/\sqrt(Z)$. Thus for large
systems, a vanishing fraction of partisans is necessary to put the
system in deadlock and prevent consensus from happening. Such finding
is consistent with our results. However, since we consider that
partisans are biased but can change state, our model shows that, in
the presence of noise, the population can actually reach consensus
when there are large fractions of partisans.

\subsection{Effect of distrust }

Because partisanship appears to remove a population's ability to reach
consensus on the correct answer, we next investigate possible ways to
counter the effect of partisan agents. A plausible strategy for
non-partisan agents is to ``discount'' the signal of partisan
agents. We thus define a discount parameter $d\in[0,1]$ with which
non-partisans weigh the information held by partisan agents. As
demonstrated by a recent study~\cite{westen06}, we must also enable
partisan agents to discount the signals of both non-partisan and
opposing partisan agents with {\em at least} the same discount rate.

Surprisingly, we find that the increased distrust among agents
actually has a deleterious effect on the efficiency of the system in
solving the density classification task (Fig. 5).  The reason for this
apparently counter-intuitive result is that $s_j=s>0$ for partisans,
so that when $d \leq 0.5$, partisan agents discount the answer of
non-partisan agents to such an extent that they will never abandon
their preferred answer. In fact, only for $d=0$ is a system
containing partisan agents able to efficiently complete the density
classification task.

\section{Discussion}

Common experience demonstrates the existence of partisanship within
groups of any kind.  One possible interpretation for partisanship is a
strong {\it a priori\/} belief that a certain answer is correct.
Alternatively, one may consider the case where agents have personal
interests that may in fact differ from the ``common good.''  In this
case individual decision rules, such as the degree of partisanship,
can be interpreted as the solution to a maximization problem at the
individual level. We model this question by assigning to a set of
rational agents an idiosyncratic utility function that each agent
tries to maximize, while the rest of the agents use the naive
strategy. The interesting case is the one in which the agent's
self-interest comes from answer ``$-1$'' being adopted while the
``common good'' comes from ``$1$'' being adopted. A utility function
for the rational agents is:
\begin{eqnarray}
  U_j &\equiv& I_j \frac{N_-}{N} + (1-I_j) E \nonumber \\
  &= &\left( 1 - \frac{3I_j}{2}\right) E + \frac{I_j}{2}\,,
\label{eqn:8}
\end{eqnarray}
where $I_j$ is the degree to which the agent values his own
``self-interest over the common good.'' If $I> 2/3$, the agent's
utility is maximized by minimizing $E$. Thus, if an agent has $I> 2/3$
the optimal individual strategy is partisanship with $s =1$. If $I \le
2/3$, then the optimal strategy depends on the strength of the noise
(Fig. 4); the stronger the noise, the more conservative the agents
should be. 

These findings thus provide a possible explanation for the emergence
of conservatism and partisanship as mechanisms to maximize individual
rather than collective advantage. The question then arises of how one
can reconcile the advantage of self-interest with the evolution toward
cooperative societies. The answer is that for individual decisions
to serve common good, societies must develop and adopt norms that
regulate self-interest~\cite{axelrod86,axelrod97}. Importantly, only
in the presence of norms and the ``metanorms'' that support their
enforcement~\cite{axelrod86}, will individuals adopt strategies that
lead toward cooperation and better social outcomes.

Our findings for the effect of partisan agents on the efficiency with
which the system completes the density classification task are
striking.  Even a small fraction of partisans can completely erase the
efficiency of the system.  It is not difficult to envision the
consequences of this result on our daily lives.  Democratic societies
face many situations in which ``difficult'' decisions must be made
~\cite{cullen05,greenstock05,white05,newhouse04,levitt05,bock01,szala05,holden05,herrera05}. Moreover,
the ability of policy makers to reach timely decisions on difficult
matters clearly increases when a strategy has broad support. Reaching
such broad consensus, unfortunately, is unlikely to occur if partisan
agents are present. Sadly, partisanship is the rational individual
strategy if there are no norms against it.

%%%%%%%%%%%%%%%%%%%%%%%%%%%%%%%%%%%%%%%%%%%%%%%%%%%
%%%%%%%%%%%%%% ACKNOWLEDGMENTS %%%%%%%%%%%%%%%%%%%%
%%%%%%%%%%%%%%%%%%%%%%%%%%%%%%%%%%%%%%%%%%%%%%%%%%%

\begin{acknowledgments}
We thank Roger Guimer\`a for discussions. M.S.-P. acknowledges the
support of CTSA grant 1 UL1 RR025741 from NCRR/NIHNIH and of NSF
SciSIP 0830338 award. L.A.N.A. gratefully acknowledges the support of
NSF awards SBE 0624318 and SciSIP 0830338.
\end{acknowledgments}

%%%%%%%%%%%%%%%%%%%%%%%%%%%%%%%%%%%%%%%%%%%%%%%%%%%%%%%%%%%%
%%%%%%%%%%%%%% MATERIALS AND METHODS--Appendices %%%%%%%%%%%
%%%%%%%%%%%%%%%%%%%%%%%%%%%%%%%%%%%%%%%%%%%%%%%%%%%%%%%%%%%%

%\section*{Methods}
%\subsection*{Efficiency of System. }

\appendix*
\section{Appendix A}

For concreteness, and without loss of generality, we assign a $57\%$
probability to state ``$1$'' in the initial configuration of the
system. By setting $p=0.57$, we avoid finite size effects (Supporting
Online Material).  We then define the instantaneous efficiency of the
coordination process as
\begin{equation}
\varepsilon(t) \equiv \frac{N_{+}(t) - N_{-}(t)}{N},
\label{eq.eff}
\end{equation}
\noindent
where $N_{+}$ is the number of agents that are in state ``$1$'' and
$N_{-}$ is the number of agents that are in state ``$-1$''.  For each
realization, we allow the system to evolve for $2N$ time steps. In
order to ensure that the strategies used by the agents are scalable,
we let the system evolve for a number of time steps proportional to
the number of agents in the system $N$.  We define the efficiency
$\overline{\varepsilon}$ of a single realization as
\noindent
\begin{equation}
  \overline{\varepsilon} \equiv \frac{1}{\tau}\sum_{t=2N-\tau}^{2N} \varepsilon(t)\,
\label{eq.effbar}
\end{equation}
\noindent
setting $\tau = N/4$. The efficiency $E$ for a given set of parameter
values is the average of $\overline{\varepsilon}$ over 1000
realizations. Crutchfield and Mitchell~\cite{crutchfield95} allowed
the system to converge to a point where all the agents have the same
state, and a realization is considered to be successful provided the
converged state is the same as the majority state. Instead of
requiring that the system reaches consensus, we focus instead on the
steady state configuration reached by the system.

%%%%%%%%%%%%%%%%%%%%%%%%%%%%%%%%%%%%%%%%%%%%%%%%%%%%%%%%%%%%
%%%%%%%%%%%%%% FIGURES %%%%%%%%5%%%%%%%%%%%%%%%%%%%%%%%%%%%%
%%%%%%%%%%%%%%%%%%%%%%%%%%%%%%%%%%%%%%%%%%%%%%%%%%%%%%%%%%%%

\clearpage
%%%%%%%%%%%%%%%%%%%%%%%%%%%%%%%%%%%%%%%%%%%%%%%%%%%%%%%%%%%%%%%%%%%%%%%
%%%%%%%%%%%%%%%%%%%%%%%%%%%%%%%%%%%%%%%%%%%%%%%%%%%%%%%%%%%%%%%%%%%%%%%

\begin{figure}
%\begin{center}
\resizebox{2\columnwidth}{!}{\includegraphics{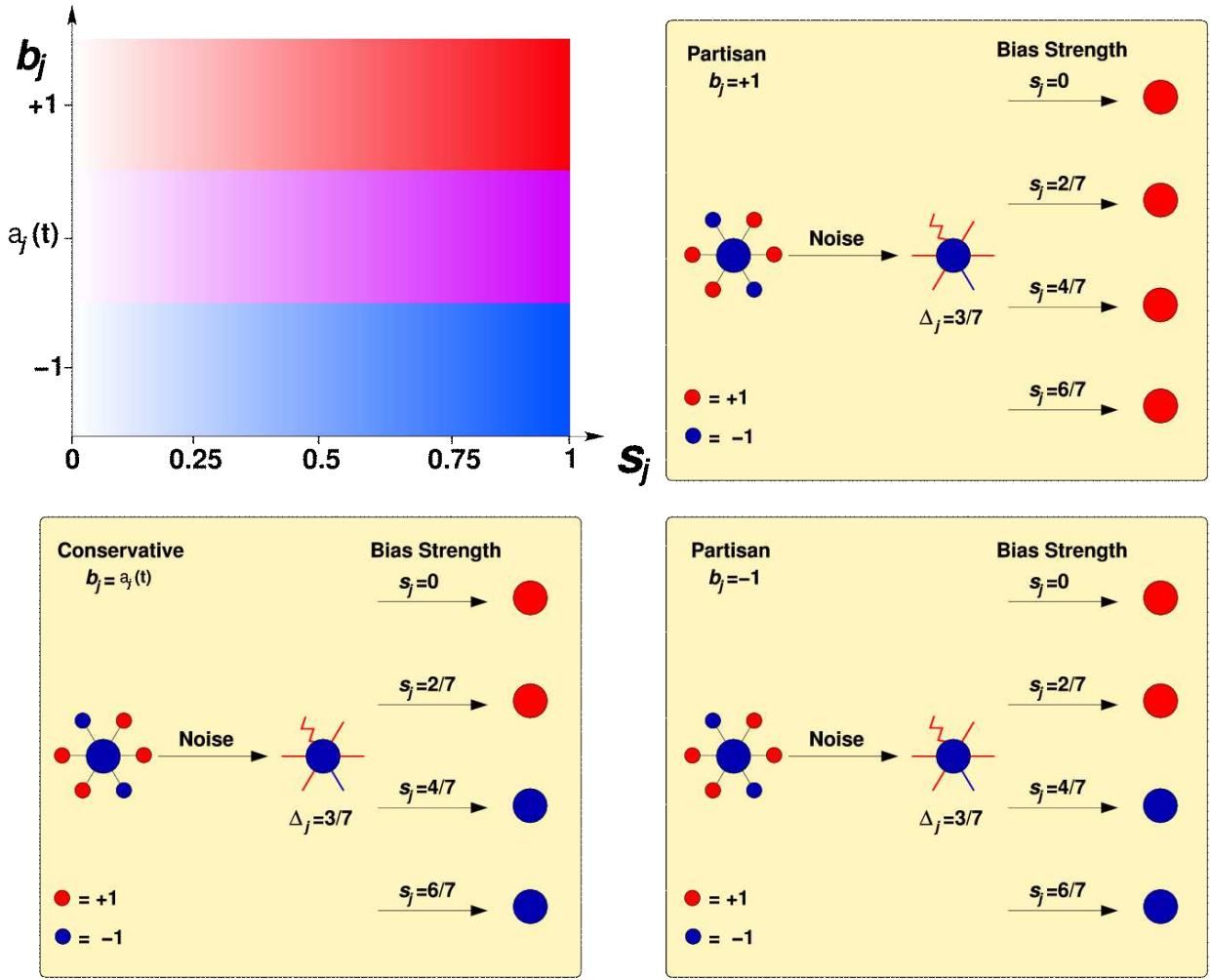}}
\caption{\baselineskip=12pt Illustration of the different agent
strategies. The strategy followed by an agent $j$ is characterized by
two parameters: $b_j$, which indicates the agent's bias, and $s_j$,
which quantifies the strength of the bias. $b_j\in \{a_j(t),-1,1\}$,
whereas $s_j \in [0,1]$. Well-intentioned agents (whether naive or
conservative) have $b_j=a_j(t)$, whereas partisan agents have $b_j=\pm
1$. If $s_j$ = 0, agents are completely trusting. As $s_j$ increases,
the level of distrust increases, so that, for $s_j$ = 1, agents will
freeze once they attain their preferred state.
}
\label{f.colors}
%\end{center}
\end{figure}

\clearpage
%%%%%%%%%%%%%%%%%%%%%%%%%%%%%%%%%%%%%%%%%%%%%%%%%%%%%%%%%%%%%%%%%%%%%%%
%%%%%%%%%%%%%%%%%%%%%%%%%%%%%%%%%%%%%%%%%%%%%%%%%%%%%%%%%%%%%%%%%%%%%%%

\begin{figure}
%\begin{center}
\resizebox{2\columnwidth}{!}{\includegraphics{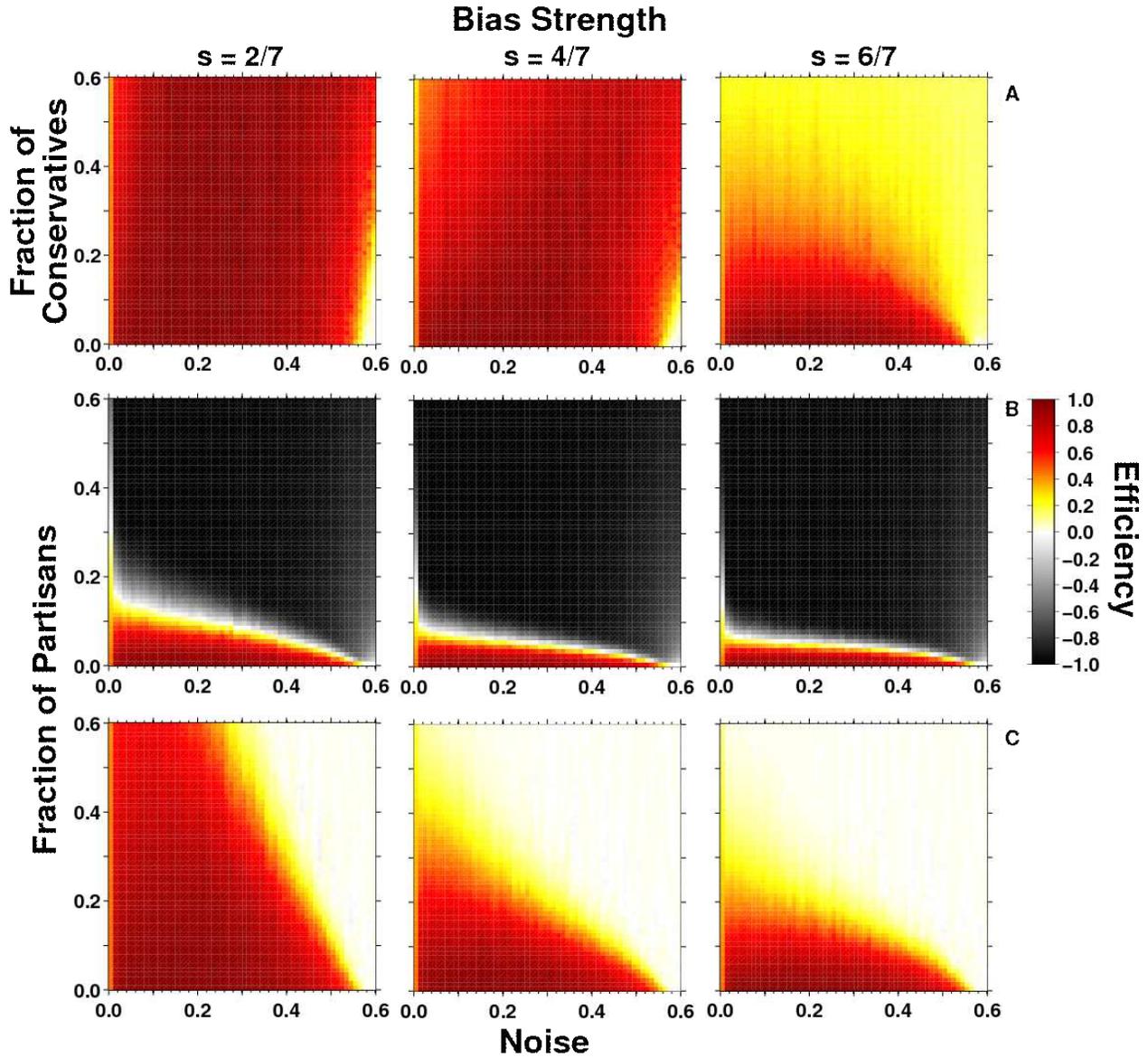}}
\caption{\baselineskip=12pt Efficiency of a population of distributed
 autonomous agents in completing the density classification task as a
 function of noise intensity and: ({\bf A}) the fraction of
 conservative agents, ({\bf B}) the fraction of partisan agents
 holding the minority state, and ({\bf C}) the total fraction of
 partisan agents in the population whose preferred states are equally
 distributed between majority and minority states. We show results
 for three bias strengths, $s=2/7$, $4/7$, and $6/7$.\\ 
 Our results suggest that (i) conservatism can be beneficial because
 it enhances the ability of the population to efficiently solve the
 density classification task for grater noise levels (panel A); (ii)
 partisanship can completely cancel the efficiency of a population in
 solving the task, even if a small fraction of partisans ($f_p\ge
 0.15$) is present (Panel B). Panel (C) shows how having partisans
 toward both answers leads to deadlock, especially at high noise
 levels for which the population as a whole will be evenly split
 between the two states.
}
\label{f.bias}
%\end{center}
\end{figure}

\clearpage
%%%%%%%%%%%%%%%%%%%%%%%%%%%%%%%%%%%%%%%%%%%%%%%%%%%%%%%%%%%%%%%%%%%%%%%
%%%%%%%%%%%%%%%%%%%%%%%%%%%%%%%%%%%%%%%%%%%%%%%%%%%%%%%%%%%%%%%%%%%%%%%

\begin{figure}
%\begin{center}
\resizebox{2\columnwidth}{!}{\includegraphics{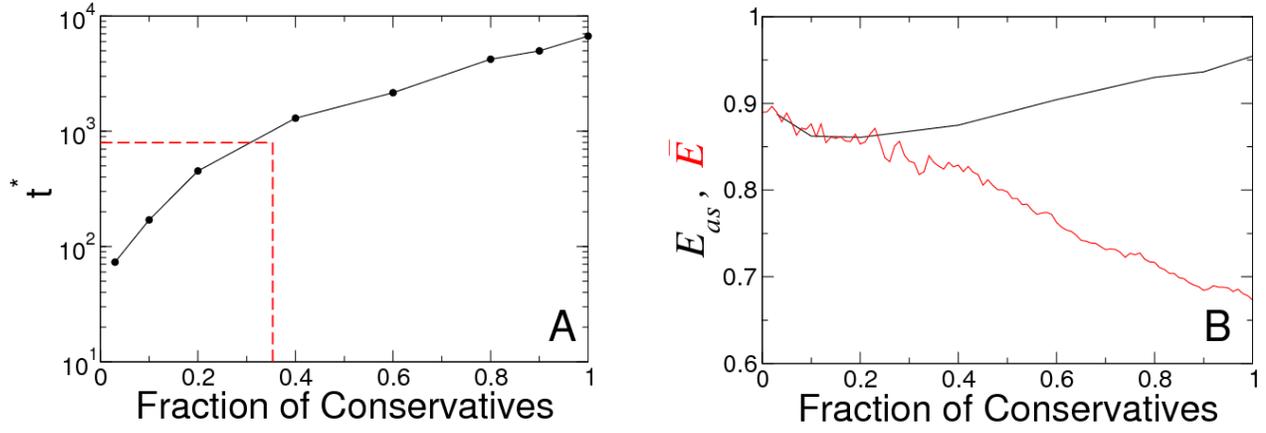}}
\caption{\baselineskip=12pt Effect of conservatism on attaining the
steady state. ({\bf A}) Time for a population to reach the stationary
state as a function of the fraction of conservatives. ({\bf B})
Comparison of the asymptotic efficiency ($E_{as}$), that is the
efficiency in the stationary state, and the efficiency $E$ attained
after $2N$ time steps as a function of the fraction of
conservatives. Note that for $f_c > 0.3$, the system cannot reach the
stationary state in the $2N$ time steps used in the simulations.}
\label{f.tc}
%\end{center}
\end{figure}

\clearpage
%%%%%%%%%%%%%%%%%%%%%%%%%%%%%%%%%%%%%%%%%%%%%%%%%%%%%%%%%%%%%%%%%%%%%%%
%%%%%%%%%%%%%%%%%%%%%%%%%%%%%%%%%%%%%%%%%%%%%%%%%%%%%%%%%%%%%%%%%%%%%%%

\begin{figure}
%\begin{center}
\resizebox{2\columnwidth}{!}{\includegraphics{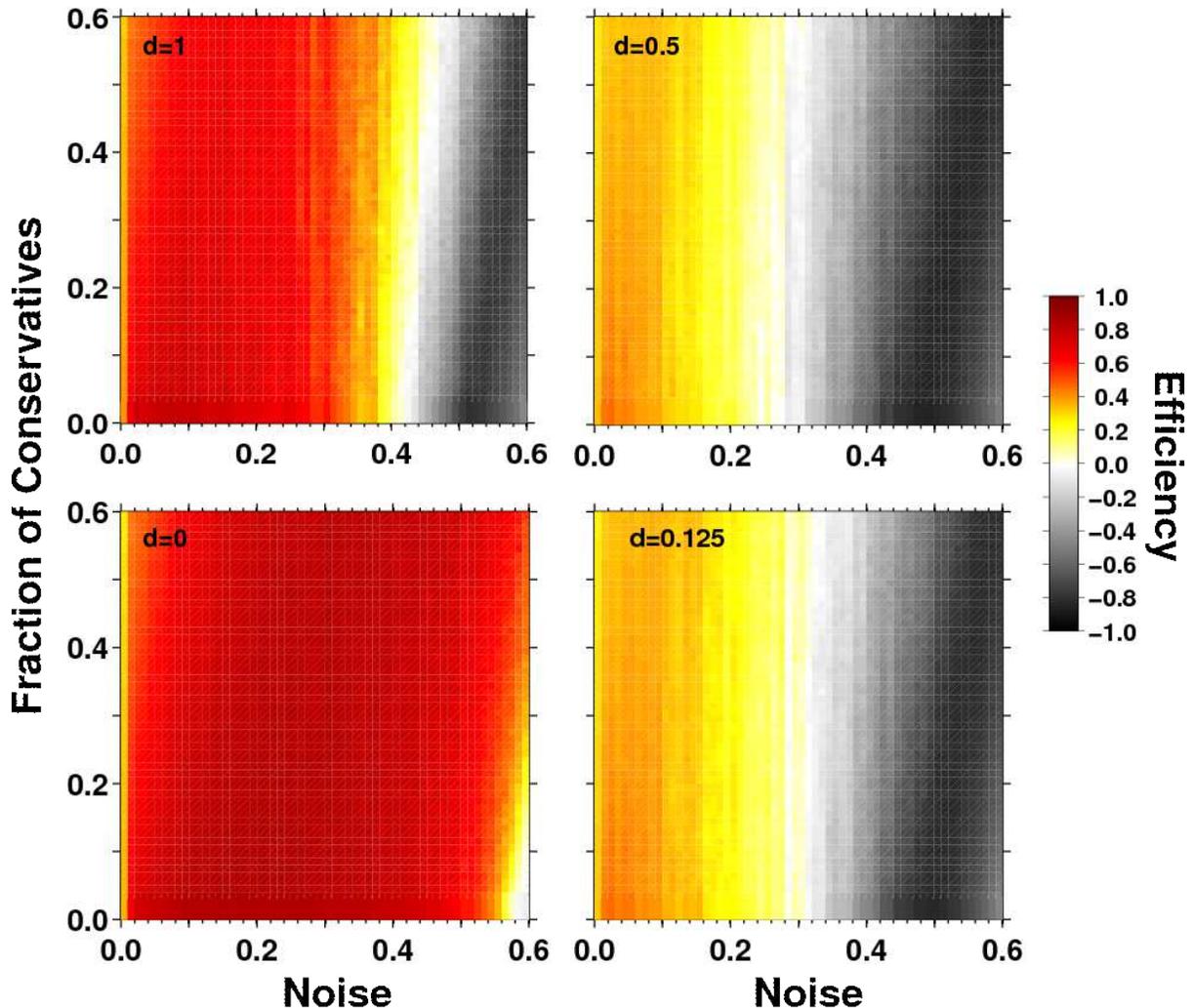}}
\caption{\baselineskip=12pt Effect of selective distrust. We consider
a system in which well-intentioned agents take partial consideration
of partisan agents' opinions and {\it vice-versa}. The discount
parameter $d$ quantifies the weight a well-intentioned agent assigns
to the opinion of a partisan agent. We consider a population with a
fraction $f_c$ of conservative agents and $5\%$ of partisan agents,
both of them with bias strength $s=2/7$, and $5\%$ of population being
partisans to the minority opinion. We show the efficiency of the
population as a function of $f_c$ and of noise. For $d > 0.5$,
partisans may converge to the positive state if a qualified majority
of their neighbors are already in that state. In such case, the
population can still attain a relatively high efficiency for a wide
range of parameter values. In contrast, when $0 < d \leq 0.5$,
partisans are unlikely to change states even when all their neighbors
are in the opposite state. In such conditions, the small fraction of
partisans acts as a constant bias toward the negative state, resulting
in a drastic reduction of the population's efficiency. Only for $d=0$,
that is when the two groups, well-intentioned and partisan agents,
completely disregard each other does the system recover the ability to
efficiently solve the density classification task.}
\label{f.discount}
%\end{center}
\end{figure}

\clearpage
%%%%%%%%%%%%%%%%%%%%%%%%%%%%%%%%%%%%%%%%%%%%%%%%%%%%%%%%%%%%%%%%%%%%%%%
%%%%%%%%%%%%%%%%%%%%%%%%%%%%%%%%%%%%%%%%%%%%%%%%%%%%%%%%%%%%%%%%%%%%%%%

\begin{figure}
\begin{center}
\resizebox{2\columnwidth}{!}{\includegraphics{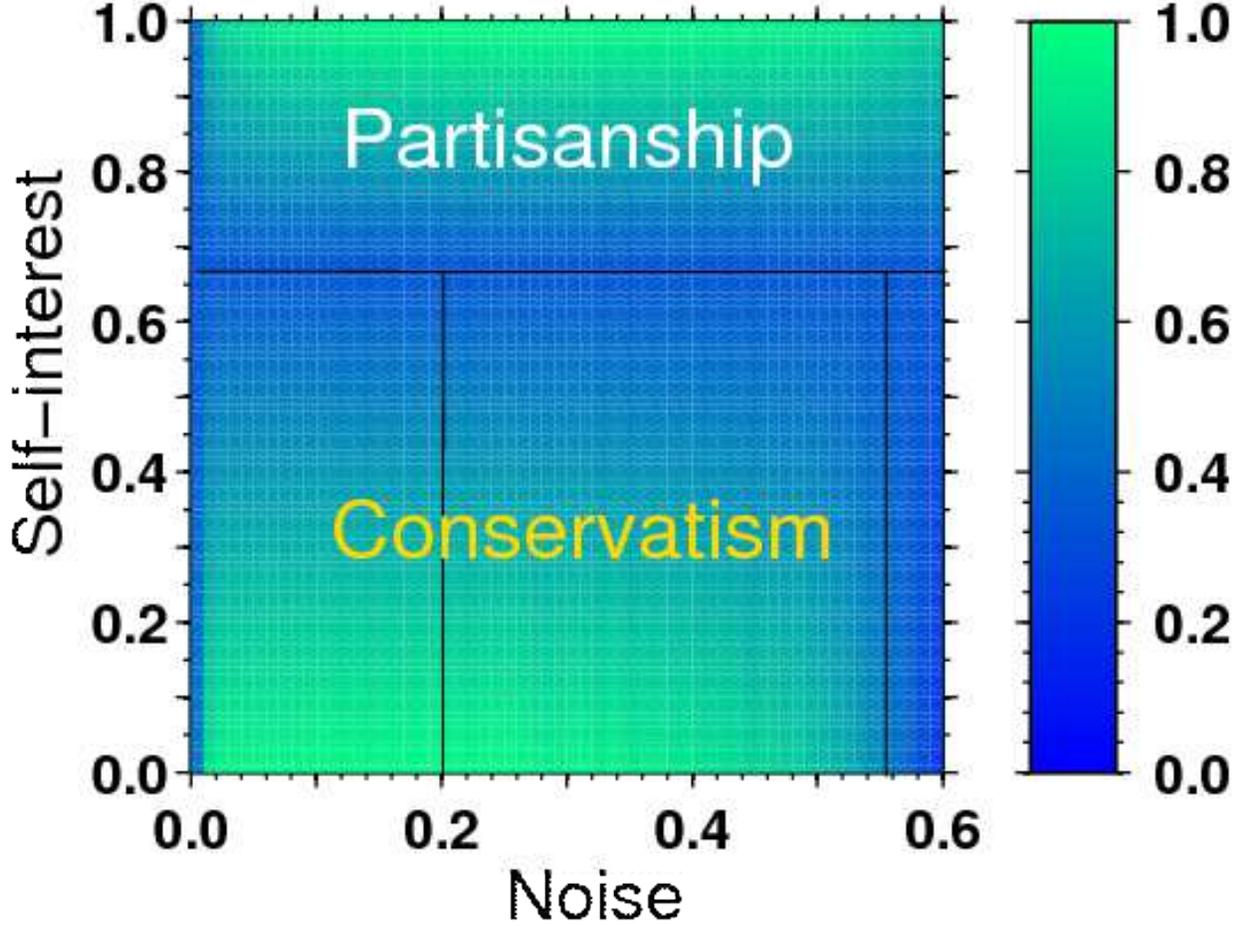}}
\caption{\baselineskip=12pt Strategy evolvability. We show the
  utility function $U$ of an agent as a function of his self-interest
  $I_j$ and the noise level $\eta$, where $U_j \equiv I_j
  \frac{N_-}{N} + (1-I_j) E^*(\eta) = \left( 1 - \frac{3I}{2}\right)
  E^*(\eta) + \frac{I}{2}$. We show results for populations with a
  10\% of non-naive agents (see Figs.~2A and 2B), that is, for each
  set of values $\{I_j,\eta\}$, we select the combination
  $\{b_j,s_j\}$ whose efficiency $E^*(\eta)$ maximizes $U$. In the
  diagram, we also show that the specific combination $\{b_j,s_j\}$
  that maximizes $U$ defines well separated regions of how different
  strategies can be optimal for different self-interest and noise
  levels. Significantly, an agent will choose to be partisan if
  $I_j>2/3$, regardless of the value of $\eta$. }
\label{f.self-interest}
\end{center}
\end{figure}

%%%%%%%%%%%%%%%%%%%%%%%%%%%%%%%%%%%%%%%%%%%%%%%%%%%
%%%%%%%%%%%%%% REFERENCES %%%%%%%%%%%%%%%%%%%%%%%%%
%%%%%%%%%%%%%%%%%%%%%%%%%%%%%%%%%%%%%%%%%%%%%%%%%%%


\begin{thebibliography}{36}
\expandafter\ifx\csname natexlab\endcsname\relax\def\natexlab#1{#1}\fi
\expandafter\ifx\csname bibnamefont\endcsname\relax
  \def\bibnamefont#1{#1}\fi
\expandafter\ifx\csname bibfnamefont\endcsname\relax
  \def\bibfnamefont#1{#1}\fi
\expandafter\ifx\csname citenamefont\endcsname\relax
  \def\citenamefont#1{#1}\fi
\expandafter\ifx\csname url\endcsname\relax
  \def\url#1{\texttt{#1}}\fi
\expandafter\ifx\csname urlprefix\endcsname\relax\def\urlprefix{URL }\fi
\providecommand{\bibinfo}[2]{#2}
\providecommand{\eprint}[2][]{\url{#2}}

\bibitem[{\citenamefont{Gigerenzer et~al.}(1999)\citenamefont{Gigerenzer, Todd,
  and the ABC Research~Group}}]{gigerenzer99}
\bibinfo{author}{\bibfnamefont{G.}~\bibnamefont{Gigerenzer}},
  \bibinfo{author}{\bibfnamefont{P.~M.} \bibnamefont{Todd}}, \bibnamefont{and}
  \bibinfo{author}{\bibnamefont{the ABC Research~Group}},
  \emph{\bibinfo{title}{Simple Heuristics That Make Us Smart}}
  (\bibinfo{publisher}{Oxford University Press}, \bibinfo{year}{1999}).

\bibitem[{\citenamefont{Gingerenzer}(2000)}]{gigerenzer00}
\bibinfo{author}{\bibfnamefont{G.}~\bibnamefont{Gingerenzer}},
  \emph{\bibinfo{title}{Adaptive Thinking: Rationality in the Real World
  (Evolution and Cognition Series)}} (\bibinfo{publisher}{Oxford University
  Press}, \bibinfo{year}{2000}).

\bibitem[{\citenamefont{Goldman et~al.}(2005)\citenamefont{Goldman, Gabriel,
  and Kaufmann}}]{goldman05}
\bibinfo{author}{\bibfnamefont{R.}~\bibnamefont{Goldman}},
  \bibinfo{author}{\bibfnamefont{R.~P.} \bibnamefont{Gabriel}},
  \bibnamefont{and} \bibinfo{author}{\bibfnamefont{M.}~\bibnamefont{Kaufmann}},
  \emph{\bibinfo{title}{Innovation Happens Elsewhere: Open Source as Business
  Strategy}} (\bibinfo{publisher}{Elsevier}, \bibinfo{year}{2005}).

\bibitem[{\citenamefont{Giere et~al.}(2005)\citenamefont{Giere, Bickle, and
  Mauldin}}]{giere05}
\bibinfo{author}{\bibfnamefont{R.~N.} \bibnamefont{Giere}},
  \bibinfo{author}{\bibfnamefont{J.}~\bibnamefont{Bickle}}, \bibnamefont{and}
  \bibinfo{author}{\bibfnamefont{R.}~\bibnamefont{Mauldin}},
  \emph{\bibinfo{title}{Understanding Scientific Reasoning}}
  (\bibinfo{publisher}{Wadsworth Publishing}, \bibinfo{year}{2005}).

\bibitem[{\citenamefont{Lesser and Storck}(2001)}]{lesser01}
\bibinfo{author}{\bibfnamefont{E.~L.} \bibnamefont{Lesser}} \bibnamefont{and}
  \bibinfo{author}{\bibfnamefont{J.}~\bibnamefont{Storck}},
  \bibinfo{journal}{IBM Syst. J.} \textbf{\bibinfo{volume}{40}},
  \bibinfo{pages}{831} (\bibinfo{year}{2001}).

\bibitem[{\citenamefont{Chesbrough}(2003)}]{chesbrough03}
\bibinfo{author}{\bibfnamefont{H.~W.} \bibnamefont{Chesbrough}},
  \bibinfo{journal}{Harv. Bus. Rev.} \textbf{\bibinfo{volume}{81}},
  \bibinfo{pages}{12} (\bibinfo{year}{2003}).

\bibitem[{\citenamefont{Page and Hong}(2001)}]{page01}
\bibinfo{author}{\bibfnamefont{S.}~\bibnamefont{Page}} \bibnamefont{and}
  \bibinfo{author}{\bibfnamefont{L.}~\bibnamefont{Hong}}, \bibinfo{journal}{J.
  Econ. Theory} \textbf{\bibinfo{volume}{97}}, \bibinfo{pages}{123}
  (\bibinfo{year}{2001}).

\bibitem[{\citenamefont{March}(1991)}]{march91}
\bibinfo{author}{\bibfnamefont{J.}~\bibnamefont{March}},
  \bibinfo{journal}{Organization Sci.} \textbf{\bibinfo{volume}{2}},
  \bibinfo{pages}{71} (\bibinfo{year}{1991}).

\bibitem[{\citenamefont{Guimer\`a et~al.}(2005)\citenamefont{Guimer\`a, Uzzi,
  Spiro, and Amaral}}]{guimera05c}
\bibinfo{author}{\bibfnamefont{R.}~\bibnamefont{Guimer\`a}},
  \bibinfo{author}{\bibfnamefont{B.}~\bibnamefont{Uzzi}},
  \bibinfo{author}{\bibfnamefont{J.}~\bibnamefont{Spiro}}, \bibnamefont{and}
  \bibinfo{author}{\bibfnamefont{L.}~\bibnamefont{Amaral}},
  \bibinfo{journal}{Science} \textbf{\bibinfo{volume}{308}},
  \bibinfo{pages}{697} (\bibinfo{year}{2005}).

\bibitem[{\citenamefont{Castellano et~al.}(2003)\citenamefont{Castellano,
  Vilone, and Vespignani}}]{castellano03}
\bibinfo{author}{\bibfnamefont{C.}~\bibnamefont{Castellano}},
  \bibinfo{author}{\bibfnamefont{D.}~\bibnamefont{Vilone}}, \bibnamefont{and}
  \bibinfo{author}{\bibfnamefont{A.}~\bibnamefont{Vespignani}},
  \bibinfo{journal}{Europhys. Lett.} \textbf{\bibinfo{volume}{63}},
  \bibinfo{pages}{153} (\bibinfo{year}{2003}).

\bibitem[{\citenamefont{Moreira et~al.}(2004)\citenamefont{Moreira, Mathur,
  Diermeier, and Amaral}}]{moreira04}
\bibinfo{author}{\bibfnamefont{A.~A.} \bibnamefont{Moreira}},
  \bibinfo{author}{\bibfnamefont{A.}~\bibnamefont{Mathur}},
  \bibinfo{author}{\bibfnamefont{D.}~\bibnamefont{Diermeier}},
  \bibnamefont{and} \bibinfo{author}{\bibfnamefont{L.~A.~N.}
  \bibnamefont{Amaral}}, \bibinfo{journal}{Proc. Natl. Acad. Sci. USA}
  \textbf{\bibinfo{volume}{101}}, \bibinfo{pages}{12083}
  (\bibinfo{year}{2004}).

\bibitem[{\citenamefont{Sood and Redner}(2005)}]{sood05}
\bibinfo{author}{\bibfnamefont{V.}~\bibnamefont{Sood}} \bibnamefont{and}
  \bibinfo{author}{\bibfnamefont{S.}~\bibnamefont{Redner}},
  \bibinfo{journal}{Phys Rev Lett} \textbf{\bibinfo{volume}{94}},
  \bibinfo{pages}{178701} (\bibinfo{year}{2005}).

\bibitem[{\citenamefont{Uzzi and Spiro}(2005)}]{uzzi05}
\bibinfo{author}{\bibfnamefont{B.}~\bibnamefont{Uzzi}} \bibnamefont{and}
  \bibinfo{author}{\bibfnamefont{J.}~\bibnamefont{Spiro}},
  \bibinfo{journal}{Am. J. Sociol.} \textbf{\bibinfo{volume}{111}},
  \bibinfo{pages}{447} (\bibinfo{year}{2005}).

\bibitem[{\citenamefont{Suchecki et~al.}(2005)\citenamefont{Suchecki,
  Egu\'{\i}luz, and Miguel}}]{suchecki05}
\bibinfo{author}{\bibfnamefont{K.}~\bibnamefont{Suchecki}},
  \bibinfo{author}{\bibfnamefont{V.~M.} \bibnamefont{Egu\'{\i}luz}},
  \bibnamefont{and} \bibinfo{author}{\bibfnamefont{M.~S.}
  \bibnamefont{Miguel}}, \bibinfo{journal}{Phys Rev E Stat Nonlin Soft Matter
  Phys} \textbf{\bibinfo{volume}{72}}, \bibinfo{pages}{036132}
  (\bibinfo{year}{2005}).

\bibitem[{\citenamefont{Crutchfield and Mitchell}(1995)}]{crutchfield95}
\bibinfo{author}{\bibfnamefont{J.~P.} \bibnamefont{Crutchfield}}
  \bibnamefont{and} \bibinfo{author}{\bibfnamefont{M.}~\bibnamefont{Mitchell}},
  \bibinfo{journal}{Proc. Natl. Acad. Sci. USA} \textbf{\bibinfo{volume}{92}},
  \bibinfo{pages}{10742} (\bibinfo{year}{1995}).

\bibitem[{\citenamefont{Tessone and Toral}(2008)}]{tessone08}
\bibinfo{author}{\bibfnamefont{C.~J.} \bibnamefont{Tessone}} \bibnamefont{and}
  \bibinfo{author}{\bibfnamefont{R.}~\bibnamefont{Toral}}
  (\bibinfo{year}{2008}), \eprint{0808.0522}.

\bibitem[{\citenamefont{Girvan and Newman}(2002)}]{girvan02}
\bibinfo{author}{\bibfnamefont{M.}~\bibnamefont{Girvan}} \bibnamefont{and}
  \bibinfo{author}{\bibfnamefont{M.~E.~J.} \bibnamefont{Newman}},
  \bibinfo{journal}{Proc. Natl. Acad. Sci. USA} \textbf{\bibinfo{volume}{99}},
  \bibinfo{pages}{7821} (\bibinfo{year}{2002}).

\bibitem[{\citenamefont{Guimer\`a et~al.}(2003)\citenamefont{Guimer\`a, Danon,
  D\'{\i}az-Guilera, Giralt, and Arenas}}]{guimera03}
\bibinfo{author}{\bibfnamefont{R.}~\bibnamefont{Guimer\`a}},
  \bibinfo{author}{\bibfnamefont{L.}~\bibnamefont{Danon}},
  \bibinfo{author}{\bibfnamefont{A.}~\bibnamefont{D\'{\i}az-Guilera}},
  \bibinfo{author}{\bibfnamefont{F.}~\bibnamefont{Giralt}}, \bibnamefont{and}
  \bibinfo{author}{\bibfnamefont{A.}~\bibnamefont{Arenas}},
  \bibinfo{journal}{Phys. Rev. E} \textbf{\bibinfo{volume}{68}},
  \bibinfo{pages}{art. no. 065103} (\bibinfo{year}{2003}).

\bibitem[{\citenamefont{Newman and Girvan}(2004)}]{newman04b}
\bibinfo{author}{\bibfnamefont{M.~E.~J.} \bibnamefont{Newman}}
  \bibnamefont{and} \bibinfo{author}{\bibfnamefont{M.}~\bibnamefont{Girvan}},
  \bibinfo{journal}{Phys. Rev. E} \textbf{\bibinfo{volume}{69}},
  \bibinfo{pages}{art. no. 026113} (\bibinfo{year}{2004}).

\bibitem[{\citenamefont{Watts and Strogatz}(1998)}]{watts98}
\bibinfo{author}{\bibfnamefont{D.~J.} \bibnamefont{Watts}} \bibnamefont{and}
  \bibinfo{author}{\bibfnamefont{S.~H.} \bibnamefont{Strogatz}},
  \bibinfo{journal}{Nature} \textbf{\bibinfo{volume}{393}},
  \bibinfo{pages}{440} (\bibinfo{year}{1998}).

\bibitem[{\citenamefont{Watts}(1999)}]{watts99}
\bibinfo{author}{\bibfnamefont{D.~J.} \bibnamefont{Watts}},
  \emph{\bibinfo{title}{Small Worlds: The Dynamics of Networks between Order
  and Randomness}} (\bibinfo{publisher}{Princeton University Press},
  \bibinfo{year}{1999}).

\bibitem[{\citenamefont{Hastie and Kameda}(2005)}]{hastie05}
\bibinfo{author}{\bibfnamefont{R.}~\bibnamefont{Hastie}} \bibnamefont{and}
  \bibinfo{author}{\bibfnamefont{T.}~\bibnamefont{Kameda}},
  \bibinfo{journal}{Psychological Rev.} \textbf{\bibinfo{volume}{112}},
  \bibinfo{pages}{494} (\bibinfo{year}{2005}).

\bibitem[{\citenamefont{Mobilia}(2003)}]{mobilia03}
\bibinfo{author}{\bibfnamefont{M.}~\bibnamefont{Mobilia}},
  \bibinfo{journal}{Phys. Rev. Lett.} \textbf{\bibinfo{volume}{91}},
  \bibinfo{pages}{028701} (\bibinfo{year}{2003}).

\bibitem[{\citenamefont{Mobilia et~al.}(2007)\citenamefont{Mobilia, Petersen,
  and Redner}}]{mobilia07}
\bibinfo{author}{\bibfnamefont{M.}~\bibnamefont{Mobilia}},
  \bibinfo{author}{\bibfnamefont{A.}~\bibnamefont{Petersen}}, \bibnamefont{and}
  \bibinfo{author}{\bibfnamefont{S.}~\bibnamefont{Redner}},
  \bibinfo{journal}{J. Stat. Mech.: Theor. Exp.} p. \bibinfo{pages}{P08029}
  (\bibinfo{year}{2007}).

\bibitem[{\citenamefont{Westen et~al.}(2006)\citenamefont{Westen, Kilts,
  Blagov, Harenski, and Hammann}}]{westen06}
\bibinfo{author}{\bibfnamefont{D.}~\bibnamefont{Westen}},
  \bibinfo{author}{\bibfnamefont{C.}~\bibnamefont{Kilts}},
  \bibinfo{author}{\bibfnamefont{P.}~\bibnamefont{Blagov}},
  \bibinfo{author}{\bibfnamefont{K.}~\bibnamefont{Harenski}}, \bibnamefont{and}
  \bibinfo{author}{\bibfnamefont{S.}~\bibnamefont{Hammann}},
  \bibinfo{journal}{J. Cognitive Neuroscience.} \textbf{\bibinfo{volume}{18}},
  \bibinfo{pages}{1947} (\bibinfo{year}{2006}).

\bibitem[{\citenamefont{Axelrod}(1986)}]{axelrod86}
\bibinfo{author}{\bibfnamefont{R.}~\bibnamefont{Axelrod}},
  \bibinfo{journal}{Am. Pol. Sci. Rev.} \textbf{\bibinfo{volume}{80}},
  \bibinfo{pages}{1095} (\bibinfo{year}{1986}).

\bibitem[{\citenamefont{Axelrod}(1997)}]{axelrod97}
\bibinfo{author}{\bibfnamefont{R.}~\bibnamefont{Axelrod}},
  \emph{\bibinfo{title}{The Complexity of Cooperation: Agent-Based Models of
  Competition and Collaboration}} (\bibinfo{publisher}{Princeton University
  Press, Princeton, New Jersey}, \bibinfo{year}{1997}).

\bibitem[{\citenamefont{Cullen and Hodgson}(2005)}]{cullen05}
\bibinfo{author}{\bibfnamefont{M.}~\bibnamefont{Cullen}} \bibnamefont{and}
  \bibinfo{author}{\bibfnamefont{P.}~\bibnamefont{Hodgson}},
  \emph{\bibinfo{title}{{\em Implementing the Carbon Tax - A Government
  Consultation Paper}}} (\bibinfo{year}{2005}).

\bibitem[{\citenamefont{Greenstock}(2005)}]{greenstock05}
\bibinfo{author}{\bibfnamefont{J.}~\bibnamefont{Greenstock}},
  \emph{\bibinfo{title}{The Cost of War}} (\bibinfo{publisher}{PublicAffairs},
  \bibinfo{year}{2005}).

\bibitem[{\citenamefont{White}(2005)}]{white05}
\bibinfo{author}{\bibfnamefont{M.}~\bibnamefont{White}},
  \emph{\bibinfo{title}{The Fruits of War: How Military Conflict Accelerates
  Technology}} (\bibinfo{publisher}{Gardners Books}, \bibinfo{year}{2005}).

\bibitem[{\citenamefont{Newhouse}(2004)}]{newhouse04}
\bibinfo{author}{\bibfnamefont{J.~P.} \bibnamefont{Newhouse}},
  \emph{\bibinfo{title}{Pricing the Priceless: A Health Care Conundrum}}
  (\bibinfo{publisher}{MIT Press}, \bibinfo{year}{2004}).

\bibitem[{\citenamefont{Levitt and Dubner}(2005)}]{levitt05}
\bibinfo{author}{\bibfnamefont{S.~D.} \bibnamefont{Levitt}} \bibnamefont{and}
  \bibinfo{author}{\bibfnamefont{S.~J.} \bibnamefont{Dubner}},
  \emph{\bibinfo{title}{Freakonomics: A Rogue Economist Explores the Hidden
  Side of Everything}} (\bibinfo{publisher}{W. Morrow}, \bibinfo{year}{2005}).

\bibitem[{\citenamefont{Bock}(2001)}]{bock01}
\bibinfo{author}{\bibfnamefont{A.~W.} \bibnamefont{Bock}},
  \emph{\bibinfo{title}{Waiting to Inhale: The Politics of Medical Marijuana}}
  (\bibinfo{publisher}{Seven Locks Press}, \bibinfo{year}{2001}).

\bibitem[{\citenamefont{Szalavitz}(2005)}]{szala05}
\bibinfo{author}{\bibfnamefont{M.}~\bibnamefont{Szalavitz}},
  \bibinfo{journal}{New Scientist} \textbf{\bibinfo{volume}{2509}},
  \bibinfo{pages}{37} (\bibinfo{year}{2005}).

\bibitem[{\citenamefont{Holden}(2005)}]{holden05}
\bibinfo{author}{\bibfnamefont{C.}~\bibnamefont{Holden}},
  \bibinfo{journal}{Science} \textbf{\bibinfo{volume}{308}},
  \bibinfo{pages}{1388} (\bibinfo{year}{2005}).

\bibitem[{\citenamefont{Herrera}(2005)}]{herrera05}
\bibinfo{author}{\bibfnamefont{S.}~\bibnamefont{Herrera}},
  \bibinfo{journal}{Nat. Biotech.} \textbf{\bibinfo{volume}{23}},
  \bibinfo{pages}{775} (\bibinfo{year}{2005}).

\end{thebibliography}
\end{document}